\documentclass[superscriptaddress,aps,amsmath,amssymb,floatfix,reprint,raggedbottom,longbibliography]{revtex4-1}
\usepackage[utf8]{inputenc}
\DeclareUnicodeCharacter{2009}{\,} 
\usepackage{mathtools}
\usepackage{tikz}
\usepackage{tabularx} 
\usepackage{array} 
\usepackage{mathtools}
\usepackage{tikz}
\usetikzlibrary{tikzmark}
\newcommand{\Larrow}{\tikz[baseline=-0.5ex]{\draw[->] (0,1ex) -- (0,0) -- (1.5ex,0);}}

\begin{document}
\title{Double-Free-Layer Stochastic Magnetic Tunnel Junctions \\  with Synthetic Antiferromagnets}
\par
\author{Kemal Selcuk}
\affiliation{\mbox{Department of Electrical and Computer Engineering, University of California, Santa Barbara, CA, 93106, USA}} 

\author{Shun Kanai}
\affiliation{\mbox{Research Institute of Electrical Communication, Tohoku University, 2-1-1 Katahira, Aoba-ku, Sendai 980-8577, Japan}}
\affiliation{\mbox{Graduate School of Engineering, Tohoku University, 6-6 Aramaki Aza Aoba, Aoba-ku, Sendai 980-8579, Japan}}
\affiliation{WPI Advanced Institute for Materials Research (WPI-AIMR), Tohoku University, 2-1-1 Katahira, Aoba-ku, Sendai 980-8577, Japan}
\affiliation{Center for Science and Innovation in Spintronics (CSIS), Tohoku University, 2-1-1 Katahira, Aoba-ku, Sendai 980-8577, Japan}
\affiliation{\mbox{PRESTO, Japan Science and Technology Agency (JST), Kawaguchi 332-0012, Japan}}
\affiliation{Division for the Establishment of Frontier Sciences of Organization for Advanced Studies at Tohoku University, Tohoku University, Sendai 980-8577, Japan}
\affiliation{\mbox{National Institutes for Quantum Science and Technology, Takasaki 370-1207, Japan}}

\author{Rikuto Ota}
\affiliation{\mbox{Research Institute of Electrical Communication, Tohoku University, 2-1-1 Katahira, Aoba-ku, Sendai 980-8577, Japan}}
\affiliation{\mbox{Graduate School of Engineering, Tohoku University, 6-6 Aramaki Aza Aoba, Aoba-ku, Sendai 980-0845, Japan}}

\author{Hideo Ohno}
\affiliation{\mbox{Research Institute of Electrical Communication, Tohoku University, 2-1-1 Katahira, Aoba-ku, Sendai 980-8577, Japan}}
\affiliation{WPI Advanced Institute for Materials Research (WPI-AIMR), Tohoku University, 2-1-1 Katahira, Aoba-ku, Sendai 980-8577, Japan}
\affiliation{Center for Science and Innovation in Spintronics (CSIS), Tohoku University, 2-1-1 Katahira, Aoba-ku, Sendai 980-8577, Japan}
\affiliation{Center for Innovative Integrated Electronic Systems (CIES), Tohoku University, 468-1 Aramaki Aza Aoba, Aoba-ku, Sendai 980-0845, Japan}

\author{Shunsuke Fukami}
\affiliation{\mbox{Research Institute of Electrical Communication, Tohoku University, 2-1-1 Katahira, Aoba-ku, Sendai 980-8577, Japan}}
\affiliation{\mbox{Graduate School of Engineering, Tohoku University, 6-6 Aramaki Aza Aoba, Aoba-ku, Sendai 980-0845, Japan}}
\affiliation{WPI Advanced Institute for Materials Research (WPI-AIMR), Tohoku University, 2-1-1 Katahira, Aoba-ku, Sendai 980-8577, Japan}
\affiliation{Center for Science and Innovation in Spintronics (CSIS), Tohoku University, 2-1-1 Katahira, Aoba-ku, Sendai 980-8577, Japan}
\affiliation{Center for Innovative Integrated Electronic Systems (CIES), Tohoku University, 468-1 Aramaki Aza Aoba, Aoba-ku, Sendai 980-0845, Japan}
\affiliation{\mbox{Inamori Research Institute of Science (InaRIS), Kyoto 600-8411, Japan}}
\author{Kerem Y. Camsari}
\affiliation{\mbox{Department of Electrical and Computer Engineering, University of California, Santa Barbara, CA, 93106, USA}}

\begin{abstract}
Stochastic magnetic tunnel junctions (sMTJ) using low-barrier nanomagnets have shown promise as fast, energy-efficient, and scalable building blocks for probabilistic computing. Despite recent experimental and theoretical progress, sMTJs exhibiting the ideal characteristics necessary for probabilistic bits (p-bit) are still lacking. Ideally, the sMTJs should have (a) voltage bias independence preventing read disturbance (b) uniform randomness in the magnetization angle between the free layers, and (c) fast fluctuations without requiring external magnetic fields while being robust to magnetic field perturbations. Here, we propose a new design satisfying all of these requirements, using double-free-layer sMTJs with synthetic antiferromagnets (SAF). We evaluate the proposed sMTJ design with experimentally benchmarked spin-circuit models accounting for transport physics, coupled with the stochastic Landau-Lifshitz-Gilbert equation for magnetization dynamics. We find that the use of low-barrier SAF layers reduces dipolar coupling, achieving uncorrelated fluctuations at zero-magnetic field surviving up to diameters exceeding ($D\approx 100$ nm) if the nanomagnets can be made thin enough ($\approx 1$-$2$ nm). The double-free-layer structure retains bias-independence and the circular nature of the nanomagnets provides near-uniform randomness with fast fluctuations. Combining our full sMTJ model with advanced transistor models, we estimate the energy to generate a random bit as $\approx$ 3.6 fJ, with  fluctuation rates of $\approx$ 3.3 GHz per p-bit. Our results will guide the experimental development of superior stochastic magnetic tunnel junctions for large-scale and energy-efficient probabilistic computation for problems relevant to machine learning and artificial intelligence.

\end{abstract}
\maketitle

\section{Introduction}

The slowing down of Moore's law has been driving the development of domain-specific computers in different fields. One approach is to build physics-inspired computers exploiting the natural physics of materials and devices for efficiency. A prominent emerging example of ``let physics do the computing'' \cite{feynman1982simulating,porod1997quantum} is based on probabilistic or $p$-bits \cite{camsari2017stochasticL,borders2019integer,chowdhury2023full}. The central idea in this field is to build \emph{programmable} networks of $p$-bits, or $p$-circuits, whose natural dynamical evolution leads to the solution of a problem of interest. The mathematical formulation of $p$-bits is related to the widely used Monte Carlo or Markov Chain Monte Carlo algorithms \cite{koller2009probabilistic}, casting a wide net for their application domains from random number generation \cite{vodenicarevic2017low,talatchian2021mutual}, machine learning \cite{kaiser2022hardware,sMTJ_daniels2020energy}, combinatorial optimization \cite{borders2019integer,aadit2022massively} to a subset of quantum simulation problems \cite{camsari2019scalable,chowdhury2023emulating,grimaldi_IEDM22,Chowdhury2023commphys}. 

Since their appearance as building blocks for probabilistic computation, there have been many proposals and implementations of $p$-bits in different material systems using a variety of stochastic phenomena \cite{woo2022probabilistic,liu2022probabilistic,park2022PerovskiteNick,dengVO2pbits2023,finol2023lanthanide}. A particularly promising possibility is the use of magnetic devices in the superparamagnetic regime which have recently been shown to produce $\approx$ GHz fluctuations at room temperature \cite{hayakawa2021nanosecond,safranski2021demonstration, sun2023PRBeasyplanedominant, Schnitzspan2023nanosecond}  using in-plane magnetic tunnel junctions (MTJ). Combined with the proven manufacturability of magnetic memory that has integrated billions of \emph{deterministic} MTJs with standard complementary metal-oxide semiconductor (CMOS) technology \cite{aggarwal2019demonstration,lee20191gbit}, the development of highly integrated and fast probabilistic computers is highly probable. 

Despite successful recent demonstrations of fast superparamagnetism in in-plane MTJs, following earlier work on perpendicular MTJs \cite{parks2018superparamagnetic,funatsu2022local}, ideal stochastic MTJs for \emph{circuit-level} p-bits have been difficult to realize. In particular, a three transistor/1 MTJ design (3T/1MTJ) \cite{camsari2017implementing} has been explored as a compact and energy-efficient implementation of a p-bit in discrete \cite{borders2019integer,kaiser2022hardware,grimaldi_IEDM22,kobayashi2023heterogeneous}  and integrated implementations  \cite{daniel2023experimental}. Various other possibilities beyond the 3T/1MTJ p-bit design have been considered, including a spin-orbit-torque design (proposed in Ref.~\cite{camsari2017stochasticL}) and later experimentally demonstrated in Ref.~\cite{yin2022IEDMscalableising}  and a comparator based design \cite{grimaldi_IEDM22,kobayashi2023heterogeneous}, with possibly different sMTJ requirements. Moreover, entirely different approaches to create p-bits exploiting the  stochasticity observed in the switching characteristics of \emph{deterministic} MTJs have been proposed \cite{fukushima2014spin,rehm2022stochastic,Shim2017stochasticSOT,zink2022review,mccray2020electrically}. In this paper, we focus on the necessary design requirements for the 3T/1MTJ p-bit. This design makes minimal modifications to the currently integrated 1T/1MTJ circuit in the commercialized spin-transfer torque magnetoresistive random access memory (STT-MRAM). The 3T/1MTJ p-bit does not explicitly require spin-torque control on magnetization characteristics, and its variation tolerance is greatly improved since it uses low-barrier nanomagnets. 

\begin{figure}[!t] 
\centerline{\includegraphics[width=1\linewidth]{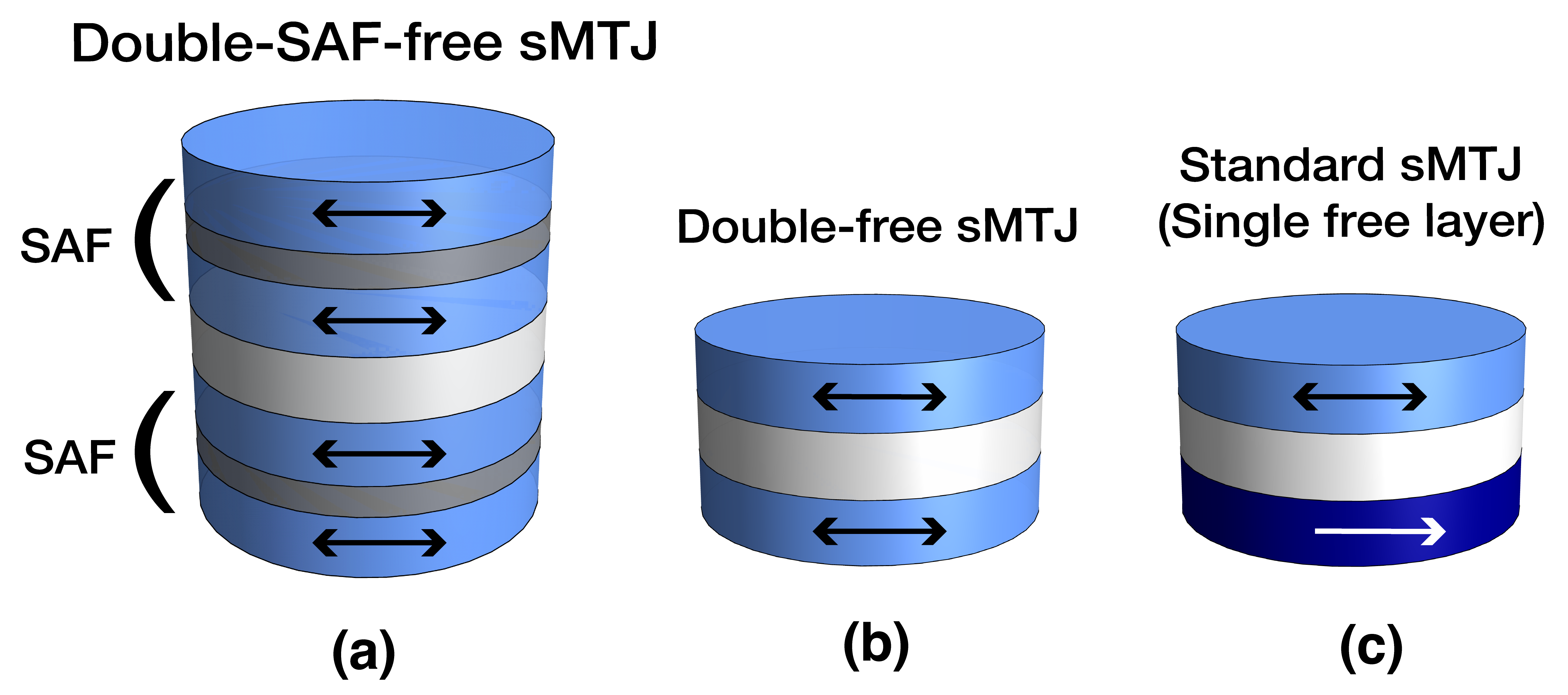}}
\caption{\textbf{Stochastic MTJ (sMTJ) designs} (a) This work: Double-SAF-free sMTJ where the low barrier free layers are replaced by magnetically inert SAF layers. (b) Double-free-layer sMTJ which no fixed layer but two low-barrier free layers. (c) Standard sMTJ with a fixed layer and a low-barrier free layer: commonly used in literature since it minimally modifies existing stable MTJs with a fixed layer to have a low-barrier free layer.}
\label{fi:fig1} 
\end{figure}

There are several requirements for scalable sMTJ-based p-bits in this 3T/1MTJ p-bit circuit. First, the sMTJ design should be magnetically soft, fluctuating without requiring external magnetic fields but electronically hard so that even at large bias currents the free layer magnetization is not pinned. Second, the sMTJs should exhibit reasonable tunneling magnetoresistance (TMR), with \emph{uniform} randomness in resistance values \cite{hassan2021quant}. Third, the sMTJ should exhibit \emph{fast} fluctuations. A previous design \cite{camsari2021double} using double-free-layers nearly satisfies all these requirements  by providing fast zero-field fluctuations and bias-independence (FIG.~\ref{fi:fig1}). The simplicity of the double-free-layer structure is appealing as opposed to asymmetrically designed typical MTJs with fixed and free layers. However, achieving zero-bias and zero-field fluctuations comes at a price: for the thermal noise to overcome the dipolar coupling between free layers the magnetic layers need to be scaled down to $\approx 20$ nm or below, imposing fabrication challenges. 

In this paper, we propose an improvement to this earlier design by considering double-free-layer sMTJs using \emph{synthetic antiferromagnetic} (SAF) free layers (FIG.~\ref{fi:fig1}). The key intuition is to use the magnetic neutrality of SAF layers. Because the SAF layers possess antiparallel nanomagnets in close proximity, the net dipolar field emanating from the SAF layers is minimal unlike single nanomagnets whose dipolar fields favor an antiparallel configuration of the two free layers, as in the case of double-free-layer sMTJs \cite{camsari2021double}. 

 In the rest of this paper, we evaluate this proposed design by the language of spin-circuits, pioneered by Brataas, Bauer and Kelly \cite{BRATAAS2006157}. Further, we employ coupled stochastic Landau-Lifshitz-Gilbert (sLLG) equations to take magnetization dynamics into account. We apply a microscopic approach to calculate the dipolar tensors between magnets and provide an approximate autocorrelation theory with analytical results. We then simulate the proposed sMTJ in conjunction with advanced transistor nodes in SPICE, estimating energy and delays to produce random bits.

\section{Spin-circuit and stochastic LLG analysis of the proposed sMTJ}

Our approach in this paper is to start from the powerful spin-circuit approach, pioneered by Brataas, Bauer and Kelly \cite{BRATAAS2006157}, that has later been converted into explicit circuit models that can be modularly simulated in SPICE-like simulators \cite{srinivasan2013modeling,manipatruni2012modeling,camsari2015modular}. These two approaches have been extensively compared and found to be exactly equivalent \cite{camsari2015modular,camsari2015modularA}. The advantage of writing down explicit SPICE-compatible $4\times 4$ conductances is the ability to combine spin the transport physics obtained from the spin-circuit approach with time-dependent stochastic LLG solvers and advanced CMOS transistor models, all in standard circuit simulators. The sLLG solvers are rigorously benchmarked against probabilistic models such as the Fokker-Planck Equation applied to nanomagnet dynamics \cite{torunbalci2018modular,kanai2021theory}. 
This approach allows us to consider each interface separately and it can be used to provide a detailed understanding of spin currents as they travel through the 4-magnet system. Technically, the use of spin-circuit theory is strictly applicable to \emph{metallic} interfaces \cite{BRATAAS2006157} and modeling \emph{tunneling} junctions requires a different treatment such as multiplying the conductance matrices \cite{camsari2014physics} to get a tunneling magnetoresistance ratio of $2 P^2/(1-P^2)$ rather than the  giant magnetoresistance ratio of $P^2/(1-P^2)$ where $P$ is the interface polarization. Nevertheless, to keep our model fine-grained and microscopic at each interface, we use the metallic spin-circuit models (as in Ref.~\cite{manipatruni2014vector}) to study the proposed double-SAF-free MTJs as a series of ferromagnet-normal metal interfaces. To match experiments, we choose the interface polarizations ($P$) appropriately, corresponding to experimental TMR values (shown in Table \ref{tab:simulation_parameters}). 

\begin{figure*}[t!]
    \centering
    \includegraphics[width=1\textwidth]{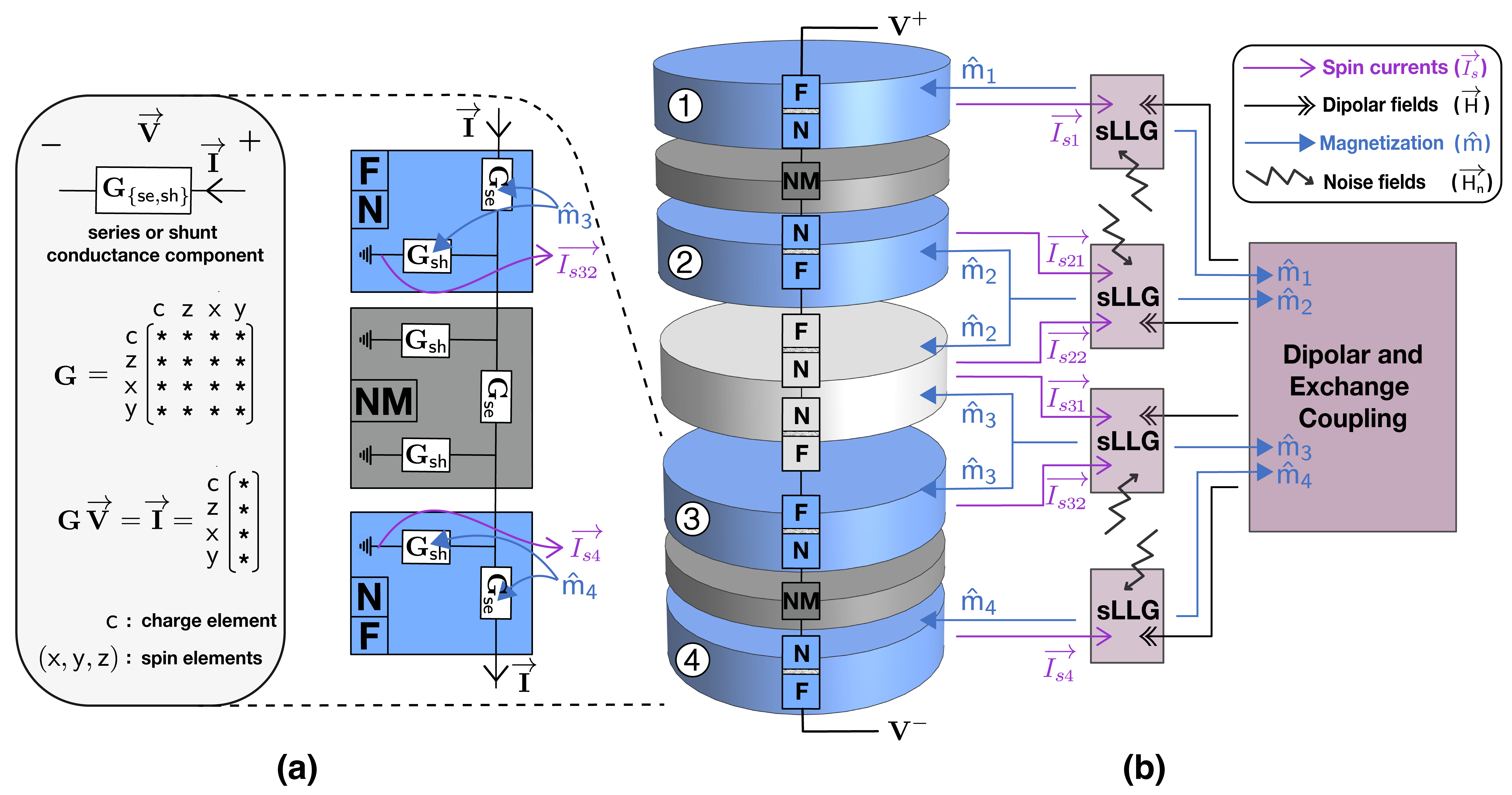}
    \caption{\textbf{Modular spin-circuit model for the double-SAF-free layer structure} (a) Close-up look on the spin-circuit transport model for a single SAF stack-layer: represented with  two ferromagnet-normal metal (F$|$N) interfaces and a normal metal (NM) block. (b) The full model coupling magnetizaton dynamics (sLLG) with spin-transport modules. The transport module produces spin-polarized currents input to sLLGs. For the two interfaces of the same magnet, we add the spin-currents incident to the magnets vectorially (e.g., $\vec{I}_{s2} = \vec{I}_{s21} + \vec{I}_{s22}$) as an input to the sLLGs, since we are in the monodomain approximation for the magnets. LLGs  in turn produce instantaneous magnetization vectors back to the interface modules. In addition, the two interfaces of the same magnet receive the same magnetization vector from their corresponding sLLG's as shown.  Dipolar, exchange and thermal noise are considered by the sLLG model. Details of the 4$\times$4 conductances and the sLLG module can be found in Ref.'s \cite{camsari2015modular,camsari2015modularA,torunbalci2018modular, nanohub:spintronics}. Numerical parameters are summarized in Table~\ref{tab:simulation_parameters}.}
    \label{fi:modularmodel}
\end{figure*}

FIG.~\ref{fi:modularmodel} shows how we decompose the full double-SAF-free layers into corresponding interfaces. F$|$N denotes a 
ferromagnet-normal metal interface, $\text{NM}$
denotes a normal, non-magnetic metal. All spin-circuits are described by 4-component currents (3 for spin and 1 for charge) related to 4$\times$4 conductance matrices (FIG.~\ref{fi:modularmodel}a). For the spacer in the SAF layers we explicitly consider the NM layers (modeling the Ruthenium) but we do not include an explicit NM layer in between the free layers for the MgO. In our full model (FIG.~\ref{fi:modularmodel}) coupling transport and magnetism, we parameterize the F$|$N interfaces with  instantaneous magnetization vectors obtained from the sLLG equations. We assume that the transport timescales are much faster than magnetization dynamics, hence we assume a lumped circuit model for each magnetization vector \cite{torunbalci2018modular}. The details of the 4$\times$4 conductance models and the sLLG solver we used in this paper have been explained in detail in Ref.'s \cite{camsari2015modular,camsari2015modularA,torunbalci2018modular}. 

\begin{figure*}[t!]
    \centering
    \includegraphics[width=1\textwidth]{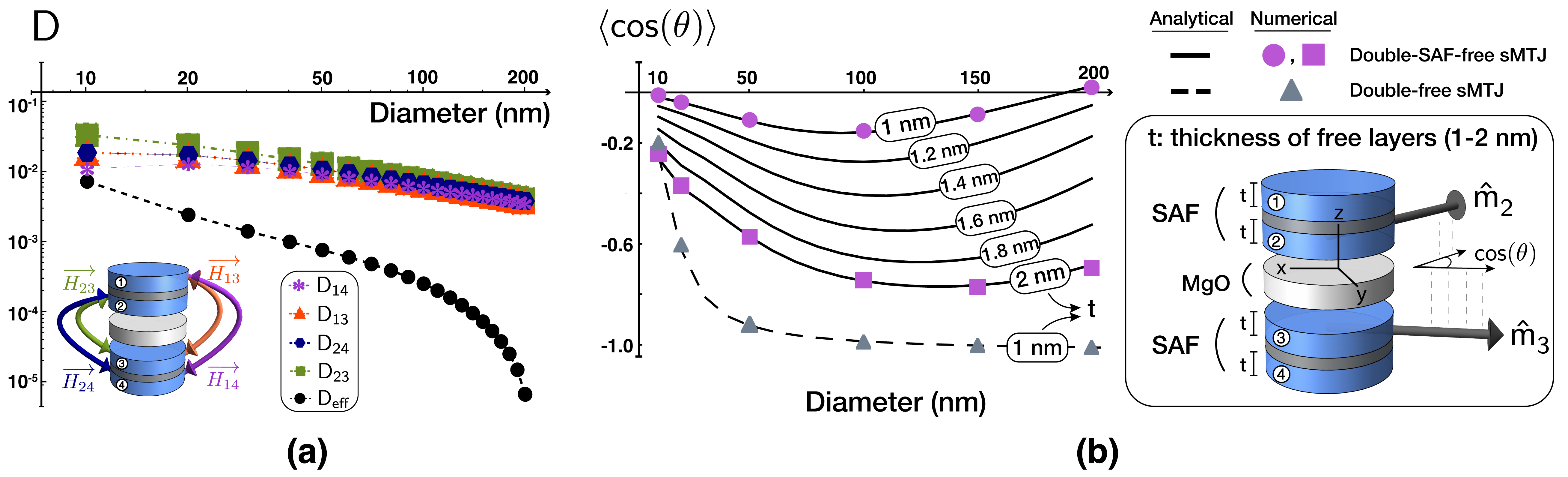}
    \vspace{-10pt}
\caption{\textbf{Dipolar coupling in the double-SAF-free layer stucture} (a) Dipolar coupling coefficients $\mathrm{D_{ij}}$ between magnets $(i,j)$ are shown at varying diameters (1 nm thickness), numerically calculated based on the approach described in Ref.~\cite{camsari2021double}.  The effective dipolar ($\text{D}_\text{eff}$) coupling (Eq.~\ref{eq:deff}) is also shown. $\mathrm{D_{12}}$ and $\mathrm{D_{34}}$ are combined with exchange tensor ($J_{ij}$) for SAF couples in the numerical model but not shown in the plot. (b) Average angle between layers (2,3), $\langle \cos \ \theta  \rangle$, are calculated numerically (markers) and analytically (solid and dashed lines), based on (Eq.~\ref{eq:bessel}), as a function of varying magnet diameters and thicknesses. Numerical calculations for the sLLG model uses a 1 ps time-step and averages are taken over 5 $\mu s$. }
    \label{fi:dipzero} 
    \vspace{-10pt}
\end{figure*}

\section{Equilibrium (Zero-bias) behavior}
In this section, we study the zero-bias behavior of the double-SAF free layer sMTJs under the influence of thermal noise, dipolar and interlayer exchange interaction fields. We first describe our magnetostatics model we use in this paper. We consider 4 identical magnets with the same volume, interacting with dipolar and exchange interactions. The magnets are assumed to be perfectly circular and zero-barrier ($k_B T \ll 1$) with easy-plane anisotropy in the  $(x-y)$ plane. Later in Section~V, we carefully analyze the validity of the easy-plane assumption. The equilibrium energy description for this system is \cite{mayergoyz2009nonlinear}: 
 \begin{align}
    E &= -2 \pi M_s^2 \text{Vol.} \Bigg(\sum_{i=1}^{n=4} \hat{m}_i^T {N}_{ii} \hat{m}_i \nonumber \\
    &\quad+ \sum_{{i,j \atop i \neq j}}^{n=4} \hat{m}_i^T ({J}_{ij}+\mathrm{{D}_{ij}}) \hat{m}_j \Bigg)
    \label{eq:Benergy}
\end{align}
\noindent where \(M_s\) represents the saturation magnetization, \(\text{Vol.}\) is the volume of each magnet, \(\hat{m}_i\) are the 3-dimensional magnetization vectors on the unit sphere, and \(N_{ii}\), and \(\mathrm{D_{ij}}\), \(J_{ij}\), are the demagnetizations, dipolar and exchange tensors, respectively. We consider exchange interactions only between SAF-pairs (1,2) and (3,4), while  dipolar interactions are considered between all $\binom{4}{2} = 6$ pairs of magnets. Throughout this paper, we assume $N_{zz}=-1$ for all magnets with all other components of the demagnetization tensor being zero. Note that this is a simplifying assumption and corrections to the demagnetization tensor \cite{Beleggia_demag} should be considered depending on magnet geometry for more detailed models. 

In Eq.~\ref{eq:Benergy}, we assume an isotropic and constant exchange model where the exchange tensor $J_{ij}$ is parameterized by a single number, i.e., $J_{ij} = J_0 \delta_{ij}$. We renormalize the exchange coupling $J_{ex}$, typically measured in units of J/$\mathrm{m^2}$ \cite{yakushiji2015perpendicular}, such that $J_{0}=J_{\mathrm{ex}}/(2\pi M_s^2 t)$, where t is the thickness of the magnets (assumed equal). In this form, $J_{ij}$ has the same units with dipolar tensors and can be directly compared. Throughout, we use a $|J_{ex}|$ value of 5 mJ/$\mathrm{m}^2$ for our numerical models, a reasonable choice considering experimental measurements \cite{yakushiji2015perpendicular,parkin_RKKYRuthenium_1991}. To calculate dipolar tensors, $\mathrm{D_{ij}}$, we follow the approach described in Ref.~\cite{camsari2021double}. The basic idea is to obtain position-dependent dipolar tensors $\rm D_{ij}(x,y,z)$ starting from the Poisson equation for the magnetic potential, $\Psi$, $\nabla^2 \Psi = \nabla \cdot \hat{m}$ to calculate dipolar tensors, and then to \emph{average} these tensors over the volume of the target magnet to obtain single $\rm D_{ij}$ values. Until the averaging step, this approach is exact. Due to the cylindrical symmetries present, the only non-zero components are  $\rm D_{xx}=D_{yy}=-D_{zz}/2$ where $(x,y)$ are the easy-plane and $z$ is the out-of-plane component.  In the rest of the paper, we drop the coordinate subscripts and use $\rm D$ for $\rm D_{xx}=D_{yy}$.

To study the equilibrium behavior of this system, we perform numerical and theoretical analysis. For our numerical results, after calculating all 6 D's for each pair of magnets and assuming a fixed interlayer exchange coupling between the SAF layers, we employ our full model in FIG.~\ref{fi:modularmodel} with 4 coupled sLLGs to obtain the average cos$ \ \theta$ between layers 2 and 3, which determines the output signal. The dipolar coefficients and the numerical average cos$\ \theta$ are shown in FIG.~\ref{fi:dipzero}a and FIG.~\ref{fi:dipzero}b, respectively. We study the average angle as a function of changing magnet thickness, t. Using a revised model without the use of SAF free layers but two FM$|$NM  layers in series, we also closely reproduce the results in Ref.~\cite{camsari2021double}. Ideally, $\langle \mathrm{cos} \ \theta \rangle$ should be zero corresponding to freely fluctuating and truly independent stochastic free layers. As shown by the triangle gray plots in FIG.~\ref{fi:dipzero}b, for the double-free sMTJ (without the use of SAFs), reaching such an independent regime requires extremely thin magnets at small diameters ($< 20$ nm). The proposed MTJ with double SAF layers on the other hand shows near independence even at large diameters for a range of ferromagnet thicknesses, indicating the magnetic inertness of the SAF layers.

\begin{figure*}[t!]
    \centering
    \includegraphics[width=1\textwidth]{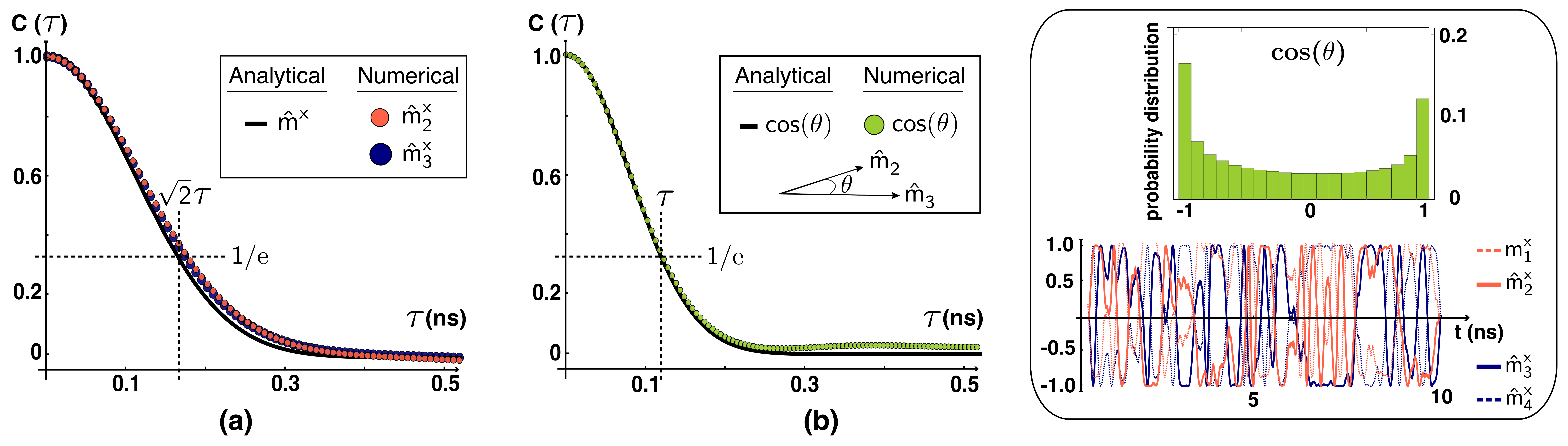}
    \caption{\textbf{Autocorrelation of magnetizations} (a) shows numerical and theoretical autocorrelation for the in-plane component of layers (2,3). (b) shows the numerical and theoretical autocorrelation of $\cos \theta $. Numerical results are obtained for a diameter of 50 nm and a thickness, t = 1 nm. Numerical results are obtained over 5 $\mu$s using a timestep of 1 ps in sLLG simulations. (c) Time-dependent magnetizations of all layers from the full numerical model. Histogram for the cosine of the angle between layers (2,3) $\cos  \ \theta $ demonstrating the necessary near uniform randomness between -1 and +1 for p-bit operation.}
    \label{fi:autocorrfig}
\end{figure*}

    Next, we provide an approximate theoretical analysis for the results shown in FIG.~\ref{fi:dipzero}. To proceed, we make simplifying assumptions. First, unlike our numerical model that assumes a finite exchange coupling, for our theoretical calculation we assume an effectively infinite exchange coupling between layers (1,2) and (3,4) that share a Ruthenium spacer between them. We note that this is simply a mathematical infinity, in practice, using experimentally achievable exchange coupling coefficients justify this assumption. This allows us to simplify the energy equation by substituting 
    $m_1^{x, y, z}$ with $-m_2^{x, y, z}$ and $m_4^{x, y, z}$ with $-m_3^{x, y, z}$ and removing constant terms from the energy as they do not change the Boltzmann probabilities: 
\begin{align}
    E &= -2 \pi M_s^2 \text{Vol.} \Big( \text{{D}}_\text{23} \hat{m}_2^T \hat{m}_3 - \text{{D}}_\text{13} \hat{m}_2^T \hat{m}_3 \nonumber \\
    &\qquad - \text{{D}}_\text{24} \hat{m}_2^T \hat{m}_3 + \text{{D}}_\text{14} \hat{m}_2^T \hat{m}_3 \Big)
    \label{eq:Benergysimpf}
\end{align}
allowing us to obtain an ``effective'' dipolar tensor between layers 2 and 3: 
    \begin{gather}
    E = -2 \pi M_s^2 \text{Vol.} \left[(\mathrm{{D}}_\text{eff}) \hat{m}_2^T \hat{m}_3\right]
    \label{eq:deff}
    \end{gather}
    where $\mathrm{D_{eff}}$=$\mathrm{D_{23}}$$-$$\mathrm{D_{13}}$$-$$\mathrm{D_{24}}$+$\mathrm{D_{14}}$. FIG.~\ref{fi:dipzero}a shows $\mathrm{D_{eff}}$ for magnets with t=$1$ nm at varying diameters, $\Phi$. We observe an approximate $\rm D_{\mathrm{eff}}\propto \Phi^{-2}$ dependence, with a decreasing effective dipolar coupling at larger diameters. These results indicate how the double SAF layer reduces the effective coupling between layers 2 and 3. Next, we use the theoretical calculation presented in Ref.~\cite{camsari2021double} where the equilibrium $\langle \cos \ \theta \rangle$ is related to the dipolar coupling coefficient D between two magnets. The idea behind this calculation is to assume a large demagnetization field and expand the Boltzmann probability, $\rho=1/Z \exp(-E/k_B T)$ around the out-of-plane component since $m_z\approx0$ for analytical approximation. Note that for this analysis, we do not set $m_z$ to 0, but keep the first order terms. With this approach, one obtains (in mks units):
 \begin{equation}
    \left\langle\cos \theta_{2,3}\right\rangle \approx \frac{I_1\left(\text{d}_\text{eff}\right)}{I_0\left(\text{d}_\text{eff}\right)}
    \label{eq:bessel}
    \end{equation}
 where \(I_n\) represents the modified Bessel function of the first kind and  \(\text{d}_\text{eff}=h_d {\text{D}_\text{eff}}\) where $h_d = \mu_0 M_s^2 \mathrm{Vol.} / k_B T$, $\mu_0$ being permeability. Using the $\mathrm{D_{eff}}$ values obtained at different diameters with Eq.~\ref{eq:bessel}, we obtain excellent agreement between the theory and our numerical model that does not assume an infinite exchange coupling between layers (1,2) and (3,4) as shown in FIG.~\ref{fi:dipzero}b. We note that the reason for the roughly constant and small $\mathrm{cos(\theta)}\approx 0-0.1 $ at small thickness (t=1 nm) can be explained by the cancellation of the approximate inverse square law dependence of the dipolar coefficient on the diameter and the square law dependence of the dipolar energy (in $h_d$) on the diameter.  The results shown in FIG.~\ref{fi:dipzero} indicate that to overcome the dipolar energy between layers (2,3), using small volume and small thickness magnets are still favorable as in double-free layer sMTJs \cite{camsari2021double}. Making CoFeB-type magnets too thin (below~$<$~2 nm)  may induce perpendicular anisotropy in  circular magnets  \cite{kobayashi2022SAF}, possibly necessitating further experimental optimization or different material systems. 

\section{Autocorrelation Theory}
    
Next, we provide an autocorrelation theory for the relative angle between magnetic layers (2,3) and compare it against our full numerical model (FIG.~\ref{fi:modularmodel}). We are interested in computing the quantity: 
\begin{equation}
C[\cos \ \theta (\tau)]=\frac{1}{T} \int_0^T \hspace{-0.2cm} \cos \ \theta (t+\tau)  \cos \ \theta (t)  dt
\label{eq:autocorr}
\end{equation}
where $\cos \ \theta$ represents the angle between layers (2,3). To simplify our analysis, to first order, we will assume that the magnets always remain in the $(x,y)$ plane and assume $m_z(t)\approx 0 $, making $\cos \ \theta (t)= m^x_2 (t) m^x_3 (t) + m^y_2 (t) m^3_y (t)$. Technically the $(x,y)$ components of magnets (2,3) are always correlated as established in FIG.~\ref{fi:dipzero}, however, at low thicknesses, they are only \emph{weakly} correlated. As such, for our theoretical analysis we make another simplifying assumption and assume that the magnets are independent of each other. This essentially means that we only need an analytical approximation for the autocorrelation of the $m_x$/$m_y$ components. Following the analysis shown in Refs.~\cite{hassan2019low,kaiser2019subnanosecond}, we proceed as follows. In circular nanomagnets, the mechanism of random fluctuations arise from the random precession of the in-plane magnetization vector around the stochastically fluctuating demagnetization field which is a function of the out-of-plane component. The autocorrelation of the in-plane magnetization vector $m_x(t)$ can be written as $C[m_x(\tau)]=\langle m_x(\tau) m_x (0) \rangle$ which reads: 

\begin{equation}
{\displaystyle \int\limits_{-\infty}^{\infty}\hspace{-5pt} \cos(\alpha m_z) \exp(-\beta m_z^2) \, dm_z} \bigg/ {\displaystyle \int\limits_{-\infty}^{\infty}\hspace{-5pt} \exp(-\beta m_z^2) \, dm_z}
\label{eq:singvect}
\end{equation}
\noindent where $\alpha$, $\beta$ are defined to be $\gamma H_d \tau$ and ${H_d M_s \text{Vol.}}/{2 
k_B T}$, respectively, while $H_d$ is the out-of-plane demagnetizaton field and the $\gamma$ is the gyromagnetic ratio,. The integral limits are extended to infinity to obtain a closed-form expression with minimal errors, due to the exponentially low probability of large $m_z$ components. This expression evaluates to, 

\begin{equation}
C[m_x(\tau)] = \exp\left(-\frac{\alpha^2}{2 \beta}\right)
\label{eq:ACFsingle}
\end{equation}

\begin{figure}[t!] 
\centerline{\includegraphics[width=1\linewidth]{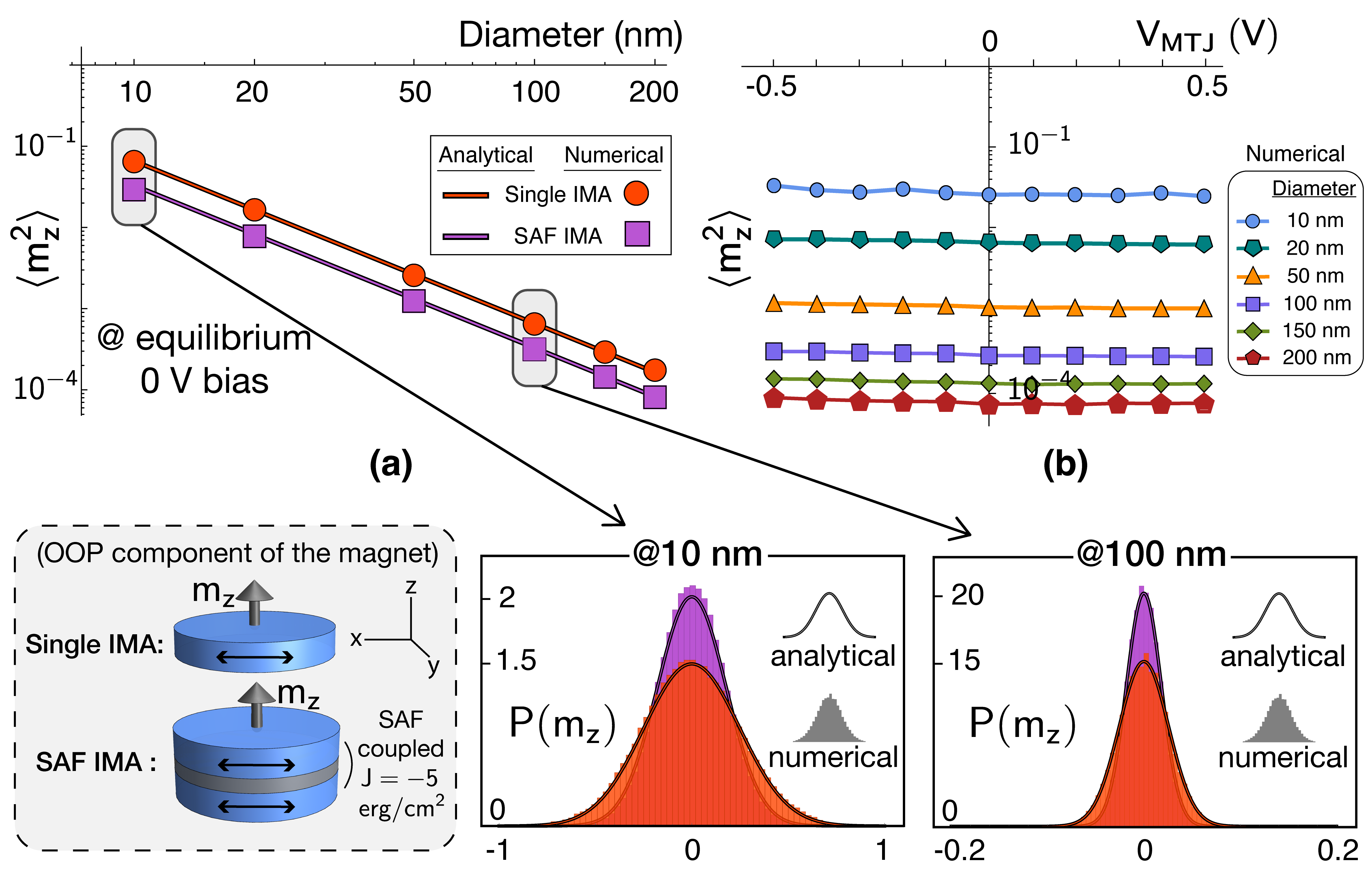}}
\caption{\textbf{Out-of-plane magnetizations of easy-plane SAFs and single nanomagnets} (a) The out-of-plane components of magnets {$m_z^2$} are evaluated as a function of diameter analytically (Eq.~\ref{eq:ima-mz} - Eq.~\ref{eq:saf-mz}) and numerically for single and SAF easy-plane magnets at zero bias (equilibrium), along with their probability distributions obtained from the Boltzmann law ($\rho\propto \exp(-E)$). (b) $\langle {m}_z^2 \rangle$ is shown numerically at different voltage biases from $-$0.5 V to 0.5 V for a range of diameters from 10 nm to 200 nm with free layer thickness of 1 nm. Each point is averaged over 250 ns  simulations with 1 ps transient time step using the .trannoise function in HSPICE. 
}
\label{fi:OOP} 
\end{figure}

\begin{figure*}[t!] 
\centerline{\includegraphics[width=1\linewidth]{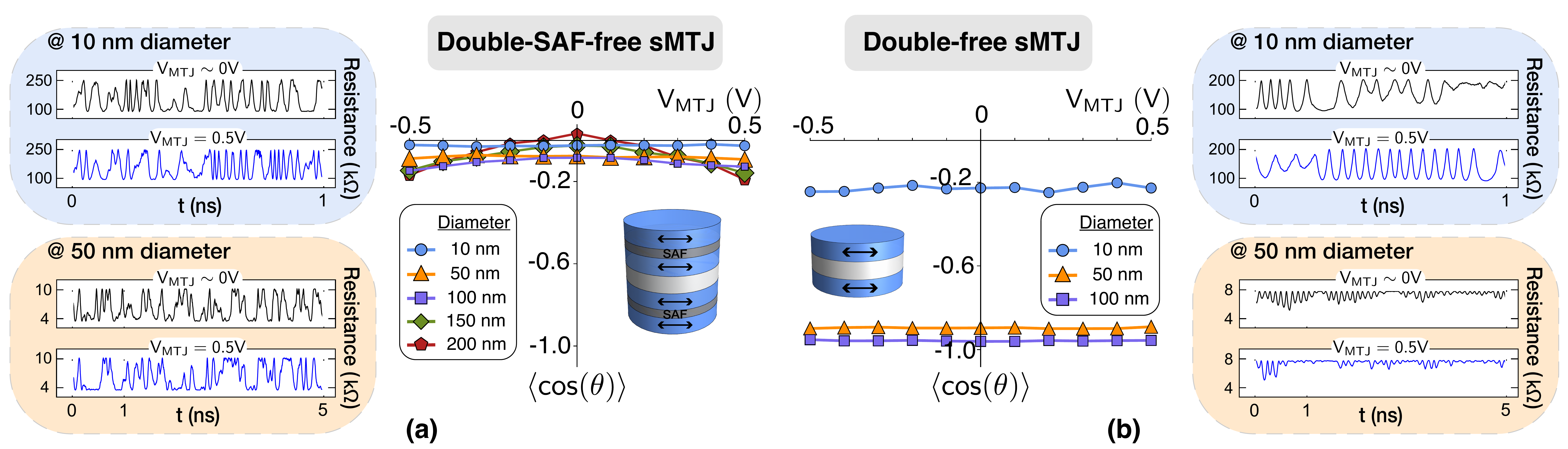}}
\caption{\textbf{Bias dependence of sMTJs} (a) Shows the time-averaged $\cos \ \theta$ as a function of the bias voltage across the MTJs for the double-SAF-free sMTJ. Left-hand side shows  corresponding resistance fluctuations at chosen diameters of 10-50 nm and bias points of 1 mV - 0.5 V. The fluctuation rates are inversely proportional to device diameter, $\Phi$, as predicted by Eq.~\ref{eq:ACFdouble}. Device exhibits  independence between the free layers, $\langle \cos \  \theta \rangle \approx 0$, with bias-independence. (b) Shows the time-averaged $\cos \ \theta$ as a function of the bias voltage for a double-free sMTJ, modeled as two F$|$N interfaces in series. Right-hand side shows corresponding resistance fluctuations at chosen diameters of 10-50 nm and bias points of 1 mV - 0.5 V. These results are in quantitative agreement with the phenomenological model introduced in Ref.~\cite{camsari2021double}. Unlike that model however we do not legislate antisymmetric currents to the magnets at the interfaces. This behavior comes out of our fine-grained interface model. Moreover, due to the uncompensated dipolar coupling the double-free sMTJ does not exhibit nearly independent fluctuations between its layers, $\langle \cos \  \theta \rangle \neq 0$. All magnets assumed to have a thickness of of 1 nm. Each data point is averaged over 1 $\mu$s with a time step of 1 ps using the .trannoise function in HSPICE.}
\label{fi:biasbehv} 
\end{figure*}

\noindent In the present context, however, we are interested in the cosine of the angle between layers 2 and 3. As such, the autocorrelation expression (assuming both magnets remain in their $(x,y)$ plane) becomes:
\begin{equation}
\frac{\displaystyle \iint \limits_{-\infty}^{+\infty} \cos[\alpha(m_{z1}-m_{z2})] \exp[{-\beta (m_{z1}^2 + m_{z2}^2)}] d\mathbf{m_z}}{\displaystyle 
\iint \limits_{-\infty}^{+\infty} \exp[{-\beta (m_{z1}^2 + m_{z2}^2)}] d\mathbf{m_z}}
\label{eq:ACFdoubleint}
\end{equation}
evaluating to,
\begin{equation}
C[\cos(\theta(\tau))] = \exp\left(-\frac{\alpha^2}{\beta}\right)
\label{eq:ACFdouble}
\end{equation}
We observe an interesting difference between the previously obtained Eq.~\ref{eq:ACFsingle} \cite{kaiser2019subnanosecond} and the new Eq.~\ref{eq:ACFdouble}. It turns out that the autocorrelation of the angle between the two independent magnets decays to $\exp(-1)$ of its value $\sqrt{2}$ times as fast as the autocorrelation of a single magnet. We corroborate this theoretical calculation with our full numerical model without any independence or in-plane rotation assumptions. We observe that at moderate diameters of 50 nm's with thicknesses of 1 nm's, the SAF layers behave as if they are nearly independent of one another. The theoretically obtained $\sqrt{2}$ factor is also observed from our numerical simulations in FIG.~\ref{fi:autocorrfig}. Faster decays of the autocorrelation on $\cos \theta $ bode well for probabilistic computing applications since they directly influence the speed with which new and independent random numbers can be obtained from sMTJ-based p-bits. 
\begin{table}[t!]
\centering
\fontsize{9}{14}\selectfont 
\caption{Simulation parameters}
\label{tab:simulation_parameters}
\begin{tabularx}{\columnwidth}{c c }
\hline
\hline
\multicolumn{1}{c}{\textbf{Parameter}} & \multicolumn{1}{c}{\textbf{Value}} \\ \hline
CoFeB/MgO/CoFeB RA-product      & 
9 $\Omega \ \mu\text{m}^2 $ \cite{lin200945nm}         \\
CoFeB/Ru/CoFeB RA-product         & $5.2 \times 10^{-3}$ $\Omega \ \mu\text{m}^2$  \cite{AKPal_1976} \cite{zhang_Ruintercnnct_2016} \\
\Larrow{} \textit{\rm corresponding   $G_0$}        & (1/RA-product)/Area $\mho$    \\
Temperature                 & 300 K        \\
$M_{s}$                         & $8\times 10^{5}$ A/m  \cite{camsari2021double}     \\
Damping Coefficient ($\alpha$)         & 0.01  \cite{sankey2008measurement}      \\
Real part of $g_{mix}$, ($G_0 a$) & a=1 \cite{camsari2015modularA,tokac_spinmixingcobalt_2015} \\   
Imaginary part of $g_{mix}$, ($G_0 b$) & b=0 \cite{camsari2015modularA} \\ 
$\lambda_{\text{sf}}$ of Ru Layer    & 14 nm  \cite{eid_magnetores_CoRu_2002}      \\
$P_{0}$ of CoFeB/MgO/CoFeB, TMR       & 0.73, TMR=115$\%$ \cite{camsari2021double} \\
$P_{0}$ of CoFeB/Ru/CoFeB, GMR        & 0.077, GMR=0.6$\%$  \cite{gubbiotti_CoRuCoGMR_2005}        \\
Diameters of sMTJs, used in p-bit        & 
$\{10, \ldots, 200\}$ nm, 25 nm    \\
Aspect ratio of Free layer               & 1  \\
Thickness of Free layer, Ru Layer   & 1-2 nm, 0.8 nm \cite{kobayashi2022SAF}    \\
Exchange Field ($J_{ex}$)             & 
$\rm  -5\times 10^{-3} \ J/m^2$ \cite{yakushiji2015perpendicular}   \\
CMOS Models                & 14 nm HP-FinFET \cite{predictive_tech}     \\
\hline
\hline
\end{tabularx}
\end{table}

\section{Out-of-plane magnetization behavior}

Our numerical results have been obtained using the simulation parameters shown in Table~\ref{tab:simulation_parameters} without making any assumptions on the out-of-plane magnetization components. Our analytical results however both for Eq.~\ref{eq:bessel} and for Eq.~\ref{eq:ACFdouble} rely on the out-of-plane magnetization components to be small. Here, we analyze the validity of this assumption both theoretically and numerically. First, consider a single easy-plane circular nanomagnet whose energy reads: 
\begin{equation}
E =  H_D M_s \mathrm{Vol.}( m_z^2)/2 
\end{equation}
where $H_D$ is the demagnetization field. From the Boltzmann law, we can compute the variance of the out-of-plane component $\langle m_z^2\rangle  = \langle\cos^2(\theta) \rangle$ from:
\begin{equation}
\langle m_z^2 \rangle = \frac{1}{Z}\int\limits_{\phi=0}^{\phi=2\pi} \hspace{-2pt} \int\limits_{\theta=0}^{\theta=\pi}\hspace{-3pt}  d\theta d\phi \cos^2(\theta) \sin(\theta)\exp(-E/k_BT)
\label{eq:bm}
\end{equation}
which evaluates to 
\begin{equation}
\langle m_z^2 \rangle = \frac{1}{{h_d}}-\frac{\displaystyle \exp{\left(-{{h_d/2}}\right)}}{\displaystyle\sqrt{\pi h_d/2} \ \mathrm{erf}\left(\displaystyle{\sqrt{{h_d/2}}}\right)}\hfill \quad \quad \quad \text{(IMA)}
\label{eq:ima-mz}
\end{equation}
where $h_d = H_D M_s \mathrm{Vol.}/k_B T$. FIG.~\ref{fi:OOP} compares Eq.~\ref{eq:ima-mz} with numerical solutions of LLG as a function of magnet diameter with excellent agreement. Interestingly, in the case of SAFs where two nanomagnets are exchange coupled, the fluctuations ($m_z^2)$ for  out of plane components of each magnet are \emph{reduced} by a factor of 2. This can be understood analytically by considering the energy of two exchanged coupled easy-plane nanomagnets:
\begin{equation}
E = \mathrm{Vol.}\left[H_D M_s\sum_{i\in \{1,2\}} \frac{\cos^2 \theta_i}{2}+ 2 J_{ex}(\hat{m}_1 \cdot \hat{m}_2) \right]
\label{eq:energyAFM}
\end{equation}
where $\cos\theta_i$ is the out-of-plane component of each easy-plane magnet.  For the experimentally relevant exchange coupling values $J_{ex}\approx -5$ erg/cm$^2$, we find that the exchange interaction dominates the demagnetization and thermal fields. In such a scenario, we are justified to assume $\hat{m}_1 \cdot \hat{m}_2 \approx -1$ in the general energy equation (Eq.~\ref{eq:energyAFM}) using the parity transformation ($\theta_2\rightarrow \pi-\theta_1$ and $\phi_1 \rightarrow \pi+\phi_2$), leading to:  
\begin{equation}
E=\left(\frac{h_d}{2} - j_{ex} + \frac{h_d}{2} \cos 2 \theta_1 \right)
\label{eq:reduced}
\end{equation}
where $j_{ex}$ is reduced to a dimensionless value similar to $h_d$.
Following a similar prescription as in Eq.~\ref{eq:bm} to compute the average out-of-plane fluctuations for SAF magnets, we get: 
\begin{equation}
\langle m_z^2\rangle =\frac{1}{2{h_d}}-\frac{\displaystyle{\exp{\left(-\displaystyle{{h_d}}\right)}}}{\displaystyle\sqrt{\pi {h_d}} \ \mathrm{erf}\left(\sqrt{{h_d}}\right)}\quad \quad \quad \quad \text{(SAF)}
\label{eq:saf-mz}
\end{equation}
Comparing Eq.~\ref{eq:saf-mz} with Eq.~\ref{eq:ima-mz}, we observe that the strong SAF coupling effectively doubles the demagnetization field and reduces fluctuations. FIG.~\ref{fi:OOP} shows that the numerical behavior is well-explained by Eq.~\ref{eq:saf-mz}. In our present context the use of SAFs as free layers compared to single easy-plane magnets \cite{camsari2021double} favorably decreases out-of-plane fluctuations, making our assumptions more reliable. Below $<20$ nm's however, one must be careful with these assumptions since large out-of-plane fluctuations may be observed. Finally, FIG.~\ref{fi:OOP}b shows the fluctuations under voltage bias conditions whose details are discussed in the next section. We observe that due to the large demagnetization fields present along with the antisymmetric magnetic configuration do not lead to instabilities (at least in the parameter ranges of interest we consider in this paper), exhibiting bias-independence on the out-of-plane fluctuations. These results are in line with earlier work where easy-plane magnets exhibit large pinning fields \cite{hassan2019low}. 

\section{Non-equilibrium (bias-dependent) behavior}
We now investigate the bias-dependence of our proposed device, using the full numerical model of FIG.~\ref{fi:modularmodel}. Considering self-consistent finite-temperature magnetization, dipolar, exchange fields and spin-circuits for transport, we numerically measure $\cos \ \theta$ at different bias voltages as shown in FIG.~\ref{fi:biasbehv}a-b. We choose experimentally-informed conductance and polarization parameters for the MTJ and Ruthenium interfaces, shown in Table~\ref{tab:simulation_parameters}. Even though we treat the SAFs layer as spin-valves composed of F$|$N interfaces, due to their relatively small interface polarizations, the main resistance modulation arises between layers 2 and 3 that are separated by the MgO layer.

    To test our fine-grained spin-circuit model that splits the system into individual F$|$N interfaces, we first design a spin-valve structure with two F$|$N interfaces to study the double-free-layer sMTJ proposed in Ref.~\cite{camsari2021double}.  Without legislating any assumptions on the spin-currents, we observe the same bias independence using our spin-circuit model with quantitatively similar results to those in Ref.~\cite{camsari2021double} which used a phenomenological model with explicitly defined antisymmetric spin-currents (FIG.~\ref{fi:biasbehv}b). This indicates that our fine-grained interface model captures detailed physics accurately. We observe that the double-free layer sMTJ shows significant correlations, measured by $\cos \ \theta$ for diameters above 50 nm's, where the magnets are completely pinned in antiparallel directions.
 
    Next, we use our full numerical model to investigate the bias-dependence of $\cos \ \theta$ for the double SAF-layer structure (FIG.~\ref{fi:biasbehv}a). We observe that for a large range of diameters $\cos \ \theta $ shows weak bias dependence and is close to zero. In addition, the double-SAF-free sMTJ should be robust to external magnetic fields, due to the magnetic insensitivity of the SAF layers \cite{kobayashi2022SAF}. We believe that our new proposed design is the most promising stochastic MTJ to date since it satisfies the three important conditions of (a) bias-independence (FIG.~\ref{fi:biasbehv}a), (b) uniform randomness in the relative magnetization angle (FIG.~\ref{fi:autocorrfig}c) and (c) fast fluctuations without requiring external magnetic fields (FIG.~\ref{fi:autocorrfig}b).     Our design improves upon earlier ones where the dipolar interaction between two closely separated nanomagnets was either ignored entirely \cite{camsari2017implementing} or required single-digit magnet diameters \cite{camsari2021double} with challenging fabrication requirements. Given the recent experimental success in observing nanosecond fluctuations in nanomagnets \cite{hayakawa2021nanosecond,safranski2021demonstration,Schnitzspan2023nanosecond}, the prospects for experimentally demonstrating the key features of double SAF sMTJs seem promising. Indeed, a recent experiment employing SAF-based free and fixed layer stochastic MTJs (albeit with different thicknesses) \cite{sun2023PRBeasyplanedominant} exhibit some of the key features of bias-independence and nanosecond fluctuations. 
    
\section{cmos circuit behavior and performance}

\begin{figure}[t!] 
\centerline{\includegraphics[width=1\linewidth]{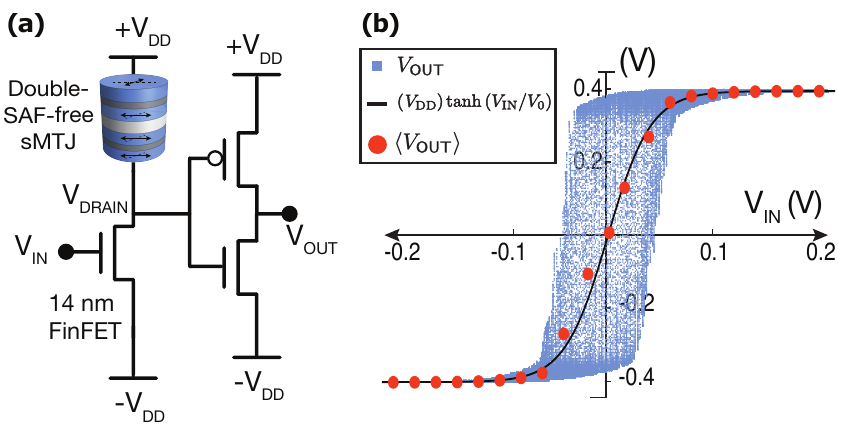}}
\caption{\textbf{3T/1MTJ p-bit circuit with double-SAF-free MTJ} (a) We use the full circuit model described in FIG.~\ref{fi:modularmodel}in conjunction with a 14 nm HP-FinFET predictive model \cite{zhao2007predictive}, simulated in HSPICE. Magnet diameters are 25 nm and thicknesses are 1 nm. The diameter is chosen to match transistor resistance. (b) The blue trace is observed by observing $\rm V_{OUT}$ as a function of $\rm V_{IN}$ while $\rm V_{IN}$ is linearly swept from $-$0.2 V to 0.2 V in a 250 ns time-window. The red dots are obtained as separate measurements where $\rm V_{IN}$ is fixed to a given value and a time-average of the output over 5 \ $\mu$s is taken with a 1 ps time step. The black line is a tanh fit to this average.}\vspace{-10pt}
\label{fi:embddpbit} 
\end{figure}

One of the main advantages of the self-consistent spin-circuit approach coupling microscopic transport physics with magnet dynamics in standard circuit simulators is in the ability to integrate them with existing CMOS models. Here, we combine our full device model presented in FIG.~\ref{fi:modularmodel} with advanced predictive technology models (PTM) for 14 nm FinFET transistors. FIG.~\ref{fi:embddpbit} shows the 3T/1MTJ topology \cite{camsari2017implementing} to design a p-bit with a binary output whose probability is controlled by an input voltage $\rm V_{IN}$. The behavioral equation for the ideal p-bit is given by: 
 \begin{equation}
    m_i=\operatorname{sgn}\left[\tanh \left(I_i\right)-\mathrm{rand_U}\right]
    \label{eq:behavioral}
    \end{equation}
For this equation to produce a smoothly varying output probability and a mean of $\langle m \rangle = \mathrm{tanh}(I_i)$, an essential requirement is that the random noise denoted by $\mathrm{rand_U}$ is \emph{uniform} and continuous between -1 and +1 \cite{hassan2021quant}. To see how the 3T/1MTJ circuit maps to Eq.~\ref{eq:behavioral}, we consider the time-varying drain voltage in FIG.~\ref{fi:embddpbit}a  given by: 
\begin{equation}
    \frac{V_\text{DRAIN}}{2 V_\text{DD}}=\frac{1}{2}-\frac{R_\text{MTJ}(t)}{R_\text{MTJ}(t)+R_\text{DS}\left(V_\text{IN})\right.}
    \label{eq:vdivider}
    \end{equation}
where $\mathrm{R_{DS}(V_{IN})}$ is the transistor resistance which varies as a function of the input voltage. Eq.~\ref{eq:vdivider} indicates another important design criterion for the 3T/1MTJ circuit to function as a p-bit, namely, for $\rm V_{DRAIN}$ to produce equal 1 and -1 fluctuations at the output of the inverter, the \emph{average} MTJ resistance needs to be matched to the transistor resistance when $\rm V_{IN}$ is at the midpoint of the sigmoid (in our circuit this is when $\rm V_{IN}=0$). When biased in its midpoint ($\rm V_{DRAIN}=0)$, the main function of the inverter is to amplify $\rm V_{DRAIN}$, which can be approximated by a sign function, making the output: 

 \begin{equation}
    \label{eq:vsimlfd}
    \begin{gathered}
    \frac{V_\text{OUT}}{V_\text{DD}} \sim \operatorname{sgn}\left[R_\text{DS}\left(V_{\text{IN}}\right)-R_\text{MTJ}(t)\right]
    \end{gathered}
\end{equation}

\noindent providing a circuit mapping to the behavioral Eq. \ref{eq:behavioral}. FIG.~\ref{fi:embddpbit} shows numerical results for this full circuit indicating how the proposed double-SAF-free MTJ approximately produces the tunable randomness necessary for the p-bit. Next, using our SPICE model we estimate several key metrics for p-bits. We delineate the power dissipation of the circuit into two branches, the sMTJ branch and the inverter branch. We obtain the power dissipation on the MTJ branch by measuring the instantaneous current flow over a 5 $\mu$s time window and multiplying this with the supply voltage, where the product is around 6.6 $\mu$W. Similarly, we measure the power dissipation on the inverter branch to be around 5.3 $\mu$W. We define a probabilistic flip based on the autocorrelation values we reported in FIG.~\ref{fi:autocorrfig} where a new flip is defined as the time when the autocorrelation roughly saturates. We estimate such an independent flip time to be around $\tau_f=0.3$ ns for the double-SAF-free MTJ. Using these numbers the energy per random number generation is estimated to be $P_{total}\times \tau_f \approx 3.6$ fJ. These values are reported in Table~\ref{tab:performance_measurements} and they are in rough agreement with those estimated in Ref.~\cite{hassan2021quant}. The important point to note however is that the proposed sMTJ does not make any simplifying assumptions about dipolar fields, the necessary uniform randomness in the MTJ or the speed of fluctuations. The total flips per second in a system of integrated p-bits, say with a million p-bits fluctuating in sparsely connected networks could lead to a system flips per second of $N/\tau_f=3.3 \times 10^{15}$ which is four to five orders of magnitude larger than what can be achieved with present-day CMOS technology \cite{aadit2022massively,romero2020high}.

\begin{table}[t!]
\centering
\fontsize{9}{14}\selectfont 
\caption{Performance measurements}
\label{tab:performance_measurements}
\begin{tabularx}{\columnwidth}{l|>{\centering\arraybackslash}X}
\hline
\hline
Power dissipation on sMTJ branch     & 6.6 $\mu$W \\
\hline
Power dissipation on Inverter branch & 5.3 $\mu$W \\
\hline
Flips per second               & 3.3 flips/ns \\
\hline
Energy per RNG                 & 3.6 fJ \\
\hline
\hline
\end{tabularx}
\end{table}

\section{Conclusion}
We proposed and analyzed a new stochastic magnetic tunnel junction with double synthetic-antiferromagnetic free layers with a modular spin-circuit model. By theoretical analysis and numerical calculations, we have shown that the proposed stochastic MTJ can meet the important requirements of bias independence, uniform randomness and fast fluctuations without any external magnetic fields. Similar to earlier double-free sMTJ designs, the structure should be relatively easy to build. The low-barrier nature of the constituent magnets should provide robustness against variations. Circuit analysis shows that the proposed device can function as a fast and energy-efficient p-bit design which can aid the development of large-scale probabilistic computers, useful for a wide variety of applications relevant to machine learning and Artificial Intelligence algorithms.\\

\section*{Acknowledgment}
KS and KYC acknowledge the U.S. National Science Foundation (NSF) support through CCF 2106260 and SAMSUNG Global Research Outreach (GRO) grant. Use was made of computational facilities purchased with funds from the National Science Foundation (CNS-1725797) and administered by the Center for Scientific Computing (CSC). The CSC is supported by the California NanoSystems Institute and the Materials Research Science and Engineering Center (MRSEC; NSF DMR 2308708) at UC Santa Barbara. SF and SK acknowledge JST-CREST JPMJCR19K3, JST-PRESTO JPMJPR21B2, and JST-AdCORP JPMJKB2305.

\end{document}